\documentclass[a4paper]{jpconf}
\usepackage{graphicx}
\begin{document}
\title{Black hole binary OJ287 as a testing platform for general relativity}

\author{M J Valtonen$^{1,2}$, A Gopakumar$^3$ , S Mikkola$^4$, K Wiik$^4$ and H J Lehto$^4$}
\address{$^1$ Finnish Centre for Astronomy with ESO, University of Turku, 21500 Piikki\"o, Finland}
\address{$^2$ Helsinki Institute of Physics, FIN-00014 University of Helsinki, Finland}
\address{$^3$ Tata Institute of Fundamental Research, Mumbai 400005, India}
\address{$^4$ Tuorla Observatory, Department of Physics and Astronomy, University of Turku, 21500 Piikki\"o, Finland}
\ead{mvaltonen2001@yahoo.com}

\begin{abstract}
The blazar OJ287 is the most promising (and the only) case  for an extragalactic binary black hole system
inspiralling under the action of gravitational radiation reaction.
At present, though it is not possible to directly observe the binary components,
it is possible to observe the jet emanating form the primary black hole. We argue that the orbital
motion of the secondary black hole is reflected in the wobble of the jet and demonstrate that 
the wobble is orbital position dependent. The erratic wobble of the jet, reported in Agudo et al. (2012),
is analyzed by taking into account the binary nature of the system and 
we find that the erratic component of jet wobble is very small.

\end{abstract}

\section{Introduction}
OJ287 is a very special case among all quasars: its optical light curve of 120 yr length shows a double periodicity of 60 yrs and 12 yrs [1]. Another remarkable feature of this quasar is that while its radiation is dominated by synchrotron radiation at most times, during the major outbursts its optical to UV spectrum is bremsstrahlung at around $3\times10^5$ K [2]. These main outbursts occur in unpolarized light, as is expected of bremsstrahlung [3]. Further,  it is the only known quasar showing bremsstrahlung outburst peaks.

Most of these facts were not known in 1995 when a detailed model of this source was constructed [4,5]. The model proposed that the major outbursts which occur at roughly 12 yr intervals are related to a 12 yr period in a black hole binary system. The detailed mechanism for generating the outbursts which occur in pairs separated by 1 to 2 years, was presumed to be impacts on the accretion disk of the primary by the secondary. Hot bubbles of plasma pulled off the accretion disk then generate the bremsstrahlung bursts when the bubbles become optically thin.
 
At present, the model has been verified in many ways [6] and subsequently, the model is 
invoked to constrain the spin of the primary black hole and to explore the possibility of testing black hole no-hair 
theorems [7,8]. It should be noted that the binary black hole model implies typical orbital velocity  $v \sim 0.1\,c $ and it explains
why higher order post-Newtonian corrections are required to explain observed major outbursts and to make predictions 
about future major outbursts. In contrast, binary pulsars have $v/c \sim 10^{-3}$, which is roughly a factor 10 higher 
than orbital velocity of the Earth.
However, it is desirable to probe further observational implications arising from the presence of binary components in the model and
this is what is pursued below.

\section{ Dynamical modeling of the wobbling radio jet in OJ287}
Supermassive black holes in active galactic nuclei typically have bi-directional jets consisting of collimated beams of matter
that are expelled from the innermost regions of their accretion disks. Further, highly variable quasars like OJ287 are 
expected to contain relativistic jets that are beamed towards the observers.
The jet from OJ287, usually observed in centimeter wavelengths, has exhibited prominent variations in its position angle in the sky (PA from here onwards) ever since the observations began in early 1980’s.
Our aim is to infer the presence of the orbital motion by following the temporal variations in the PA measurements of OJ287.
%

In Figure 1 we show the PA observations of the jet as a function of time. The data were collected by T. Savolainen and they are reported in [9]. In Figure 1 we plot also the expected variation in the PA of the jet if the jet is connected to the primary accretion disk and follows the wobble of the disk in a binary system [9]. Here the wobble is modeled by a doubly periodic sinusoidal function of time $t$ (in yr)
\begin{equation}
\Omega_1 = 1.5\times \sin(2\pi\times(t-1934.8-d)/120)
\end{equation}
\begin{equation}
\Omega_2 = 0.25\times \cos(2\pi\times(t-1934.8-d)/11.2)
\end{equation}
\begin{equation}
\Delta\Omega = \Omega_1+\Omega_2-\Omega_0.
\end{equation}
The angle $\Omega$, the nodal angle of the accretion disk, is given in degrees. We assume that the inclination $i$ of the disk is constant, since its variation was found to be much smaller than the variation of $\Omega$ in numerical simulations. The actual inclination is close to $90^{\circ}$ relative to the binary plane; in the following and in Table 1 we use the quantity which is actually $i-\pi/2$. The mean viewing angle of the jet is given by $(\Omega_0,i_0)$ which specifies the direction of the observer. The jet PA (JPA, in degrees) is then at any moment of time
\begin{equation}
\Delta\phi = -70+{\rm atan}(\tan\Delta\Omega/\sin i_0).
\end{equation}
The mean JPA $\Delta\phi=-70^{\circ}$ is obtained by fitting to the data. In actual numerical simulations the maxima of the $\Delta\Omega$ are not quite sinusoidal but more highly peaked.

There are three free parameters in fitting this function to the data: the two components of the mean viewing angle $(\Omega_0,i_0)$ of the jet, and the time delay $d$ (in yr) of the response of the jet to the variations in the disk. The results of fitting are shown in Table 1. In column 1 we indicate the data set to be compared with: TS = cm-wave data, IA = 7 mm-wave data [10] (2 yr averages) and CV = optical polarization data in one yr bins, calculated by C. Villforth and reported in [9]. Columns 2 and 3 give the parameters $i_0$ and $\Omega_0$, respectively, and column 4 the delay time $d$ in years. The last column gives the rms error of the fit. These are the best fits found for the radio jet data as well as for the polarization data. In the latter case it is assumed that the electric vector arising from radiating electrons is parallel to the jet (i.e. the magnetic field is perpendicular to the jet, presumably due to a compression in the radiating knot). The opposite case, a magnetic field parallel to the jet, was reported in [9]; it gives a considerably worse fit to the present doubly sinusoidal model.
\begin{table}
\caption{\label{1}Model fits.}
\begin{center}
\begin{tabular}{lllll}
\br
Data&$i_0$&$\Omega_0$&$d$&rms\\
\mr
TS&-1.08&1.91&16.9&7.5\\
TS&-0.60&1.96&27.7&7.7\\
TS&0.99&-1.92&78.5&7.2\\
CV&1.36&0.48&0.05&23.3\\
CV&0.90&1.07&11.2&22.2\\
IA&0.35&0.15&11.2&16.9\\
IA&0.27&0.90&22.1&14.4\\
\br
\end{tabular}
\end{center}
\end{table}

\begin{figure}
\begin{center}
\includegraphics [width=3in]{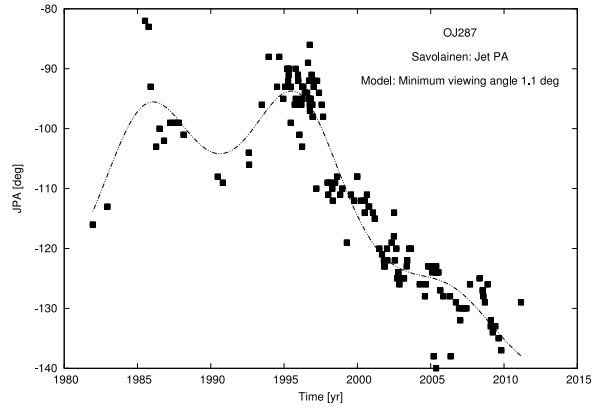} 

\end{center}
\caption{\label{label}Observations of the jet position angle in OJ287 at cm wavelengths. The line represent model described in the text.}
\end{figure}

In Figure 2 we show the average PA from the 7 mm observations by [10]. The observations at this wavelength are available only since mid 1990’s, but they have a better resolution than the cm-wave maps. Thus they refer to the jet orientation closer to the core than the cm-wave maps. An interesting feature is the major jump in the position angle of the jet around 2004. The line in the figure refers to a model where the jet responds to the changing disk, but now with a different viewing angle and time delay than in the case of the cm-wave fit.

\begin{figure}
\begin{center}
\includegraphics[width=3in]{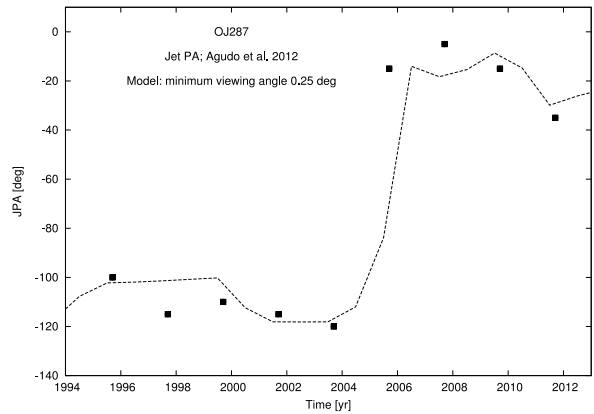}
\end{center}
\caption{\label{label}Observations of the JPA in OJ287 at 7 mm wavelength binned at 2 yr intervals. The line refers to a model calculated in [9].}
\end{figure}

\section{Interpretation}
The variations in the orientation of the accretion disk, due to the binary influence, are transmitted to the central component in about 10 yr [9]. We may assume that this variation is communicated to the jet, starting from the near jet, and proceeding outwards. We may now calculate the speed at which this happens; we call it kink speed.

In [9] it was found that the optical polarization angle responds to the disk changes with a 13 yr time delay. The present sinusoidal model suggests a 11 yr delay which is not significantly different. The source of optical emission is presumably in the jet, but closer to the central black hole than the resolved radio jet. The best fit (Figure 2) to the 7 mm data is obtained if the time delay is 22 yr, i.e. the orientation of the 7 mm jet lags behind the variations in the central component by about 11 yr. The best solution for the cm-wave jet is provided by the 79 yr time delay, i.e. its response is delayed by $\Delta t_{kink}$$\sim$ 68 years with respect to the optical core (Figure 1). A slightly worse solution is obtained with $\Delta t_{kink}$ $\sim$ 6 years.

However, since we are looking at a relativistic jet almost head-on (the viewing angle is $\sim$ $1^{\circ}$ to $2^{\circ}$), the observed time intervals $\Delta$$t$ are compressed depending on the speed of the outflow. The knots in the jet presumably take part in a flow with close to the speed of light (Lorentz factor $\Gamma_{knot}\sim20$ [11]). They take about a year to come from the central component to the observed cm-wave jet. The kink in the jet proceeds more slowly, and reaches the 7 mm-wave jet before the cm-wave jet. Thus typically, the two parts of the jet are misaligned. Our fits show that the misalignment is about 1 degree.

Let us estimate the kink speed $\beta_{kink} $ (relative to the speed of light $c$). The apparent speed of the kink is
\begin{equation}
\beta_{akink} = \beta_{kink}\sin\theta/(1-\beta_{kink}\cos\theta)
\end{equation}
while for the knot a corresponding equation is valid.

Taking the ratio
\begin{equation}
\beta_{aknot}/\beta_{akink} = \Delta t_{kink}/\Delta t_{knot} = \beta_{knot}(1-\beta_{kink}\cos\theta))/\beta_{kink}(1-\beta_{knot}\cos\theta),
\end{equation}
we find, using the observed $\Delta t_{kink}/\Delta t_{knot}\sim50$ and $\theta\sim1^{\circ}$, the Lorentz factor $\Gamma_{kink}$. From 
\begin{equation}
\beta_{kink}^2 = 1-1/\Gamma_{kink}^2,
\end{equation}
we get $\Gamma_{kink}\sim3$. For $\Delta t_{kink}/\Delta t_{knot}\sim5$ we find $\Gamma_{kink}\sim8$.
Thus the kink speed is lower than the knot speed, perhaps because the jet has a fast spine which carries the knots and a slower sheath responsible for the jet confinement.

\smallskip

\end{document}